\documentclass[pra,twocolumn,showpacs,floatfix]{revtex4}
\usepackage{amsmath}
\usepackage{amsfonts}
\usepackage{graphicx}
\usepackage[usenames]{color}
\newcommand{\beq}{\begin{equation}}
\newcommand{\eeq}{\end{equation}}
\newcommand{\beqa}{\begin{eqnarray}}
\newcommand{\eeqa}{\end{eqnarray}}
\newcommand{\ba}{\begin{array}}
\newcommand{\ea}{\end{array}}

\begin{document}

\title{Universal scaling in a trapped Fermi super-fluid in the 
BCS-unitarity crossover}

\author{S. K. Adhikari\footnote{adhikari@ift.unesp.br; URL: 
www.ift.unesp.br/users/adhikari}} \affiliation{$^1$Instituto de
 F\'{\i}sica Te\'orica, UNESP - S\~ao Paulo State University, 
 01.140-070 S\~ao Paulo, S\~ao Paulo, Brazil}

\begin{abstract} 
Using numerical simulation based on a 
density-functional equation for a trapped Fermi super-fluid valid along 
the BCS-unitarity crossover, we establish robust scaling over many orders 
of magnitude in the observables of the system as a function of fermion 
number. This scaling allows to predict the static properties of the
system, such as energy, chemical potential,  etc.,
for a large number  of fermions, over the
crossover, from the knowledge of those for a
small number ($\sim 4-10$) of fermions.
\end{abstract}
\pacs{03.75.Ss, 71.10.Ay, 67.85.Bc, 05.30.Fk}

\maketitle

\section{Introduction:} The Pauli principle among identical fermions 
leads to an effective repulsion which avoids the collapse of the 
hadronic universe.  A many-fermion system dominated by Pauli 
repulsion exhibits universal behavior with the  physical 
attraction playing a minor role. Such universal behavior manifests in 
correlation and scaling in the observables of few- and many-fermion 
systems  \cite{braaten}. This universality is prominent in the 
limit of zero Fermi-Fermi interaction in the Bardeen-Cooper-Schrieffer 
(BCS) theory of super-fluid (SF) fermions leading to universal properties of 
low-temperature superconductors \cite{bcs}, of cold neutron matter and 
neutron star \cite{baker}, and of a trapped Fermi SF 
\cite{review} at zero temperature. This universality also manifests at 
unitarity \cite{blume,CB}, 
when the atomic scattering length $a$ tends to infinity ($a 
\to \infty$). Both at the BCS limit ($a\to -0$) and unitarity, the 
scattering length ceases to be a scale of length, and the only length 
scale of the problem in these limits is $k_F^{-1}$ where $\hbar k_F$ is 
the Fermi momentum. Consequently, all energies are universal functions 
of Fermi momentum  $\hbar k_F$ (or of Fermi 
energy $E_F=\hbar^2k_F^2/2m$), where $m$ is the atomic mass, and the system
acquires universal behavior with robust scaling  involving  the stationary observables, such as 
energy,  chemical potential, etc.,  as a function of number of fermions.

How the above scaling appears in a fully paired $N$-fermion SF 
with equal number of spin up and down atoms   
confined in a spherically-symmetric trap $U({\bf r})$ can be seen in the 
local-density approximation (LDA)  \cite{review,bulgac,corr}, where the system obeys
\begin{eqnarray}
\label{lda}
U({\bf r})+\frac{\hbar^2}{2m}[3\pi^2 n({\bf r})]^{2/3}\xi  = 
\mu_0,
\end{eqnarray}
 where 
$\xi=1$ in the BCS limit and $=0.44$ (obtained \cite{th3} from 
Monte Carlo calculations on a uniform Fermi SF) at unitarity,
$n({\bf r})$ is the SF density, 
$\mu_0$ is the chemical potential, and 
$U({\bf r}) = m\omega^2 r^2/2 $ is the harmonic trap
with $\omega$ the trap frequency and normalization 
$\int n({\bf r})d {\bf r}=N$. The second term on the left-hand
side of Eq. (\ref{lda}) is the chemical potential of a uniform 
SF Fermi gas. 
  The total 
chemical potential $\mu=\mu_0 N$ and energy $E$ of the trapṕed 
system are given by
\begin{eqnarray}
\label{mu}
\mu = N\int d{\bf r}\left[ U({\bf r}) + \frac{\hbar^2}{2m}(3\pi^2 n)^{2/3}\xi \right]n({\bf r}),\\
E= N \int d{\bf r}\left[ U({\bf r}) +\frac{3}{5}\frac{\hbar^2}{2m}(3\pi^2 n)^{2/3}\xi \right] n({\bf r}).
\label{energy}
\end{eqnarray}
 It is useful to work in terms of dimensionless variables in harmonic 
oscillator units, obtained by setting $\hbar =m=\omega=1$, when 
lengths are expressed in units of $l=\sqrt{\hbar/(m\omega)}$ and 
energy in $\hbar\omega$. The normalization condition for the density 
$n$  leads to $\sqrt{n(r)}=[2 (3N)^{1/3}\sqrt \xi
-r^2]^{3/2}/(3\pi^2)$ 
for $2 (3N)^{1/3}\sqrt\xi >r^2$ and 0 otherwise. 
Using this in Eqs. (\ref{mu})
and (\ref{energy}) we get $\mu =(3N)^{4/3}\sqrt \xi /3$, 
$E =(3N)^{4/3}\sqrt \xi /4.$ An extended LDA (ELDA) was suggested \cite{corr} 
for this energy to include corrections for small $N$: 
$E =(3N)^{4/3}\sqrt \xi(1+\alpha/N^{2/3})/4,\, \alpha=0.5$ 
($\alpha$ chosen to provide agreement with  Monte Carlo data 
\cite{blume,CB}
for 
trapped Fermi SF). 

The energy  $E=(3N)^{4/3}/4$ in the BCS limit ($\xi=1$)
also follows from a
consideration of the $N$-fermion system 
in a spherical harmonic trap at zero temperature. 
The degeneracy of the $i$th state 
of harmonic oscillator of energy $\hbar\omega(i+3/2)$ is 
$(i+1)(i+2)/2$. If we consider many paired states up to $i=I (I>>1) $ 
completely 
full, 
then $E_F = I$ (in units of $\hbar\omega$)
and the total number of fermions, 
considering two (spin up and down) fermions in each state,
is $N=\sum_{i=0}^I 
(i+1)(i+2)=(I+1)(I+2)(I+3)/3\approx I^3/3$.  
The total energy in oscillator units 
is 
$E=\sum_{i=0}^I i
(i+1)(i+2)=
I(I+1)(I+2)(I+3)/4\approx I^4/4$. Consequently, $E_F=(3N)^{1/3},
k_F=\sqrt 2 (3N)^{1/6}$,
and $E=(3N)  ^{4/3}/4$, consistent with the LDA above. 
The energy at unitarity $E=(3N)  ^{4/3}\sqrt \xi/4$
differs from this energy at the BCS limit by a numerical factor 
to take into account 
the atomic attraction. The chemical potential, with  dimension of energy, 
differs from energy  by only a    numerical factor consistent with the 
universal behavior.

The $E$ and $\mu$ vs. $N$ scalings  valid in the BCS and 
unitarity limits are  useful in predicting $E$ and $\mu$ 
 for a large $N$  from a knowledge of those for a 
small $N$.  
In this paper we address two interesting questions related to these 
scalings. (a) Are the scalings 
dependent on the LDA? (b) Can these scalings 
valid at the BCS and unitarity limits be extended along the 
BCS-unitarity crossover \cite{1}? 

The LDA, lacking a gradient term, does not 
include properly the variation of the SF density near the surface. This 
can be remedied by, following a suggestion of von Weizs\"acker \cite{von}, 
including a proper gradient term in the LDA and thus transforming it 
into a density-funcional (DF) equation 
Corrections to the gradient term  have also been considered \cite{corr}.
The coefficient of the gradient 
term is fixed by requiring the DF equation to be Galilei invariant and 
also equivalent to the hydrodynamical flow equations \cite{LS}.  
The crossover from the   weak-coupling
BCS
limit to unitarity \cite{1}
has been an  active area of
research  \cite{blume,th3,castin}  after
the experimental realization  \cite{excross}
of  this  crossover
 in a trapped dilute Fermi SF
by manipulating an external background magnetic field
near a Feshbach resonance. 
Also, by 
modifying the bulk chemical potential term in the LDA to one appropriate 
for the BCS-unitarity crossover, the above scalings can be extended to the 
full crossover.

\section{Density-functional formulation:}

To study the universality along the crossover, we use a
Galilei-invariant DF formulation of the trapped
Fermi
SF \cite{LS,LS2,LS4}, equivalent to a
hydrodynamical model with  the correct
phase-velocity relation \cite{review} ${\bf v}=\hbar \nabla \theta
/(2m)$, where ${\bf
v}$ is the SF velocity, and $\theta$ the
phase of the order parameter at position ${\bf r}$:
\begin{eqnarray}\label{1}
&&\left[-\frac{\hbar^2}{8m}\nabla^2+U+
\mu(n,a)\right]\sqrt{n({\bf r})}=\mu_0\sqrt{n({\bf r})},\\
\label{2}
&& \mu(n,a)=\frac{\hbar^2}{2m}(3\pi^2n)^{2/3}g({n^{1/3}a}),\\
\label{3}
\label{gnew}
&&g(x)=1+\frac{(\chi_1 x-\chi_2 x^2)}{(1-\beta_1 x+\beta_2 x^2)},\\
&&\mu=N\int d {\bf r}\biggr[\frac{\hbar^2}{8m}|\nabla \sqrt{n}|^2+Un
+\mu(n,a)n\biggr],\label{4}\\
&& E=N\int d {\bf r}\biggr[\frac{\hbar^2}{8m}|\nabla \sqrt{n}|^2+Un
+\int_0^n \mu(n',a) dn'
\biggr],\label{4}
\end{eqnarray}
where $\chi_1, \chi_2, \beta_1$  and $\beta_2$ are parameters. 
If we choose 
$\chi_1 = 4\pi/(3\pi^2)^{2/3}, \chi_2=300, \beta_1=40,$
$\beta_2=\chi_2/(1-\xi)$,  
the DF equation (\ref{1}) (i) produces energies that 
 agree with the fixed-node Monte Carlo (FNMC) \cite{blume}  and
Green-function Monte
Carlo (GFMC) \cite{CB}  results for  the energies
of a trapped  Fermi SF
at   unitarity (for $N\le 30$), 
and over the crossover \cite{LS2,vs} for $N=4$ and 8,  (ii)
provides a
smooth
interpolation between the energies of a SF
at the BCS and unitarity limits \cite{LS,vs}.
 (iii) Moreover, the bulk chemical potential $\mu (n,a)$ of 
Eq. (\ref{1}) is exact 
at unitarity and  satisfies \cite{LS,LS2} the weak-coupling BCS limit  
$\lim_{x\to 0} g(x) \to 1 +4\pi x/(3\pi^2)^{2/3}$ \cite{LS3}.

The gradient term in Eq. (\ref{1})
corresponds \cite{LS} to a quantum
pressure
in the equivalent hydrodynamical equations and
provides a correction to
the
LDA \cite{bulgac}.  LDA is a good approximation for a large $N$, when the
bulk chemical potential $\mu(n,a)$ $-$ a positive term responsible for
Pauli repulsion in
the system even for attractive (negative) $a$ $-$ 
is
very large. The gradient term is consistent with the hydrodynamic flow
of paired fermions of mass $2m$ \cite{review,LS}.

\begin{figure}[t]
\includegraphics*[width=\columnwidth]{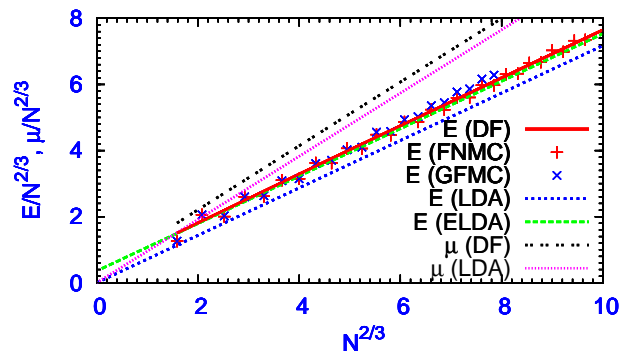}
\caption{\label{fig1}(online color at www.lphys.org)
 Energy and chemical potential of a trapped Fermi SF in oscillator units 
as a function of atom number at unitarity obtained from a 
solution of Eq. (\ref{1}) and from 
FNMC \cite{blume} and  GFMC \cite{CB}
calculations as well as LDA and ELDA calculations.
}
\end{figure}

\section{Numerical results of scaling}

We solve the DF equation (\ref{1}) by the split-step Crank-Nicolson 
method by transforming it into a time-dependent equation and using the 
FORTRAN programs provided in Ref. \cite{MA}. The space and time steps used in 
discretization of the equation were 0.05 and 0.001, respectively. 

The calculated energy and chemical potential of the trapped Fermi SF as 
a function of the number of Fermi atoms are exhibited in Fig. 
\ref{fig1}. We also compare the results for energy with those obtained 
by the FNMC \cite{blume} and  
GFMC \cite{CB} calculations as well as LDA and ELDA calculations. 
It is found that the DF and ELDA
results for energy are in better agreement with the Monte Carlo 
simulations than the LDA results.

\begin{figure}[t]
\includegraphics*[width=\columnwidth]{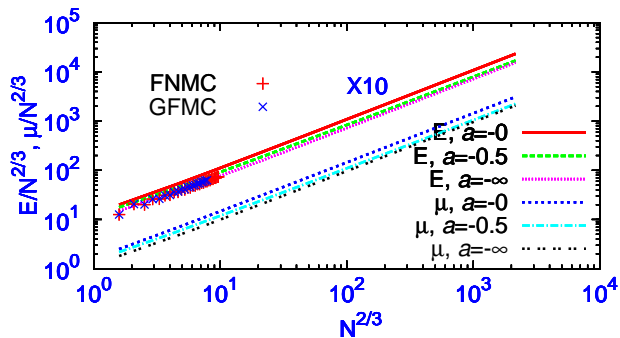}
\caption{\label{fig2}(online color at www.lphys.org)
Same as in Fig. \ref{fig1}, but with results 
for different scattering lengths $a$: $a= -0$ (BCS limit),
$a= -0.5$ (in BCS-unitarity crosover), $a=-\infty$ (unitarity).
(The LDA results are not shown.)
}
\end{figure}

An analysis of the DF results for energy and  chemical potential of Fig. 
\ref{fig1} reveals that these functions maintain the following quantities   
\begin{eqnarray}
&&\delta(N)\equiv [E(N)/N^{2/3}-0.37]/N^{2/3}, \\
&&\kappa(N)\equiv [\mu(N)/N^{2/3}-0.27]/N^{2/3},  
\end{eqnarray}
 fixed at constant values independent of $N$. This is a consequence of 
scalings  $E(N)/N^{2/3}$  and 
 $\mu(N)/N^{2/3}$ vs. $N^{2/3}$ as shown in Fig. \ref{fig1}. 
 If the functions $\delta(N)$ and $\kappa(N)$ were really universal,
then they should maintain approximate
constant values 
along the BCS-unitarity crossover independent of $N$, 
although weakly dependent on the atomic
 scattering length $a$. To demonstrate it, we plot in 
  Fig. \ref{fig2} $E(N)/N^{2/3}$ and $\mu(N)/N^{2/3}$ vs. $N^{2/3}$ as 
obtained from the DF equation (\ref{1}) along the BCS-unitarity 
crossover for three values of $a$ 
representing the BCS limit, unitarity and one  in the crossover.
Figure  \ref{fig2} clearly illustrates the robust scaling.

\begin{figure}[t]
\includegraphics*[width=\columnwidth]{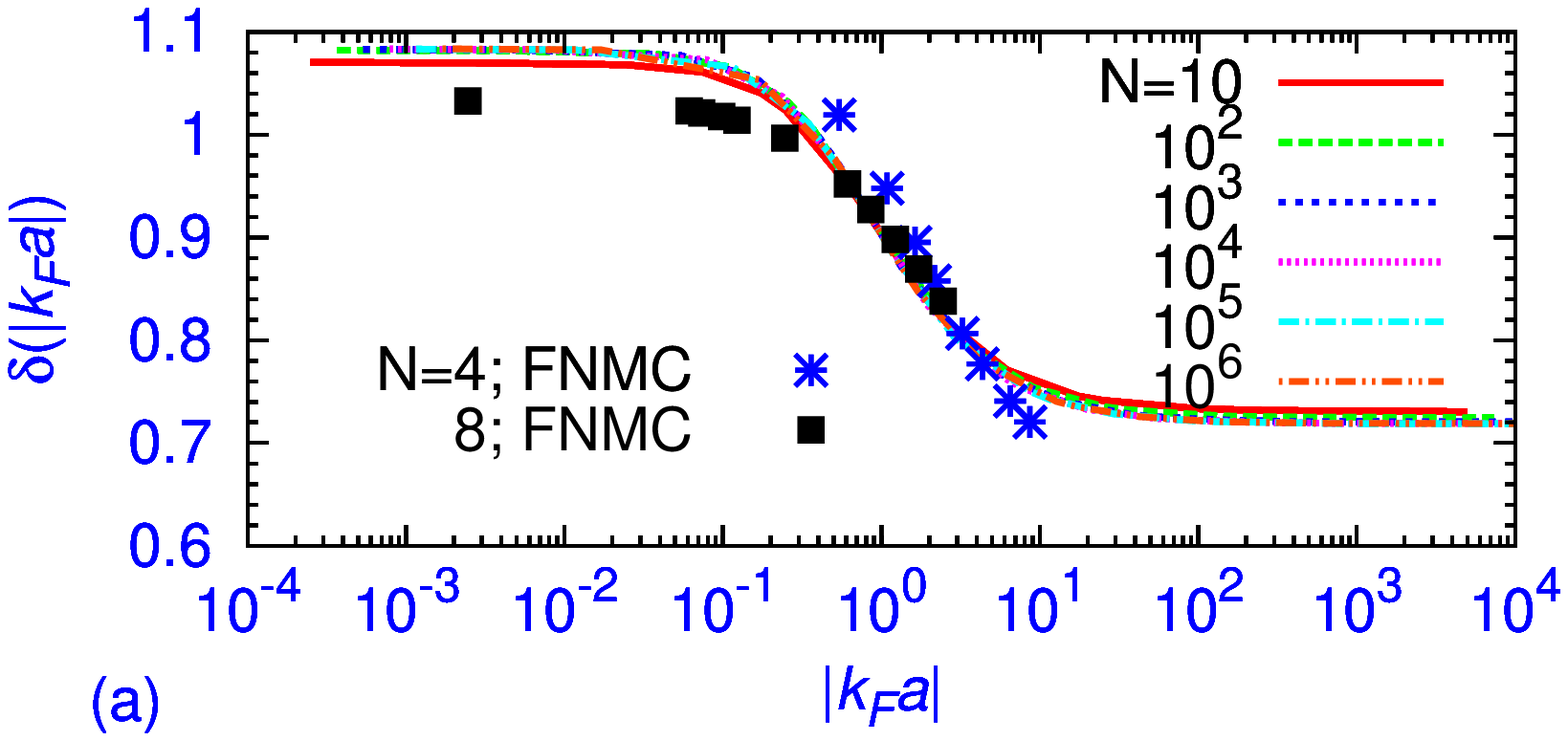}
\includegraphics*[width=\columnwidth]{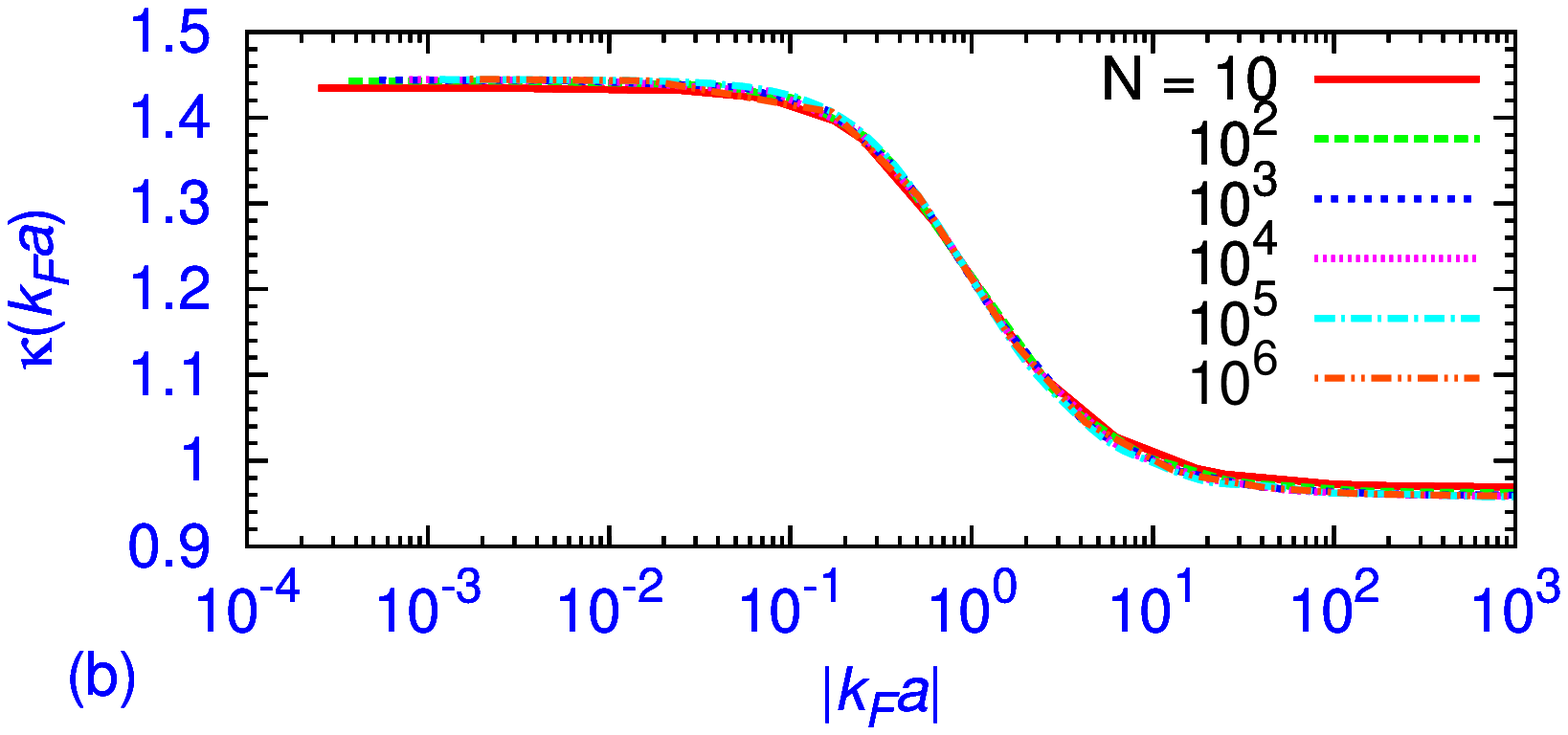}
\caption{\label{fig3}(online color at www.lphys.org)
(a) $\delta(N,k_Fa)=[E(N,k_Fa)/N^(2/3)-0.37]/N^(2/3)$ vs. $|k_Fa|$
 for different $N$ from the solution of DF equation (\ref{1}) (lines)
and from  FNMC calculations \cite{vs} (points). (b) 
$\kappa(N,k_Fa)=[\mu(N,k_Fa)/N^(2/3)-0.27]/N^(2/3)$ vs. $|k_Fa|$
 for different $N$ from the solution of DF equation (\ref{1}) (lines).
}
\end{figure}

To take the full advantage of the above scalings at unitarity, 
as shown in 
Figs. \ref{fig1} and \ref{fig2}, 
along the full BCS-unitarity crossover we plot 
$\delta(N,k_Fa)= [E(N,k_Fa)/N^{2/3}-0.37]/N^{2/3}$  and 
$\kappa(N,k_Fa)= [\mu(N,k_Fa)/N^{2/3}-0.27]/N^{2/3}$ vs.  $|k_Fa|$ in 
Figs. \ref{fig3} (a) and (b), respectively, where $k_F=\sqrt 2 
(3N)^{1/6}$. It should be noted that $k_Fa$ 
is a model-independent
dimensionless  measure of  atomic interaction and should hence be used in the 
study of universal scaling. 
If the Fermi SF were really 
dominated by the universal scaling, 
a plot of $\delta(N,k_Fa)$ and 
$\kappa(N,k_Fa)$ vs.  $|k_Fa|$ should lead to universal curves 
indepndent of 
$N$ and this is indeed so as can be seen from Figs. \ref{fig3}.

\begin{table}[!ht]
\begin{center}
\caption{Dimensionless energies $E/[\hbar\omega(3N)^{4/3}]$ of
a trapped Fermi
SF along the crossover.
The last two  columns, \{4\} and \{10\},  give the predicted
energies for $N=10^5$  employing scaling (\ref{scaling})
using the energies for
$N'=4$ and 10, respectively.  }
\label{table1}
\begin{tabular}{|r|r|r|r|r|r|}
\hline
$a\backslash N=$ & 10  & $10^2$ &
   $10^5$ &  $10^5$\{4\}&$10^5$\{10\} \\
\hline
   $-0.001$  &0.2656  & 0.2539   &  0.2499 &0.2428
&0.2470
\\
   $-0.01$  & 0.2652 & 0.2531  &  0.2459  & 0.2393
& 0.2434  \\
   $-0.1$  & 0.2550 & 0.2364  & 0.2046 &0.2027
&0.2049
\\
   $-1$  & 0.2079 & 0.1869     &   0.1714
&0.1731
&0.1740
\\
   $-10$  & 0.1897 &  0.1731  & 0.1665 &0.1686
&0.1693
\\
   $-100$  & 0.1875 & 0.1714   &  0.1659
&0.1682
&0.1688
\\
\hline
\end{tabular}
\end{center}
\end{table}

The universal nature of the energy curves as illustrated in Fig. 
\ref{fig3}
(a) allows us to predict energy of the $N$ fermion system 
from a knowledge 
of the energy of $N'$ fermions from the universal relation
\begin{equation}\label{scaling}
\frac{E(N,a)/N^{2/3}-0.37}{N^{2/3}}=\frac{E(N',a)/N'^{2/3}-0.37}{N'^{2/3}}.
\end{equation}
We have done so and our results for energy as calculated 
by solving the DF equation are exhibited in Table 
\ref{table1} \cite{LS4}, where   we display energy 
along the
crossover for different $N$ and $a$. We
also calculated the energies for $N=10^5$ atoms using the
energies for $N'= 4$ and 10. 
The predictions so obtained 
 listed in Table 1, compare well with the calculated results. 
This demonstrates 
the usefulness of the scaling (\ref{scaling}) for energy. 

\section{Summary and discussion}

 From a  numerical study of the static
properties of a
trapped Fermi SF using a Galilei-invariant DF
formulation \cite{LS}, equivalent to a generalized hydrodynamic
formulation with
the correct phase-velocity relation \cite{review},  we establish that,
because of the
dominance of the Pauli repulsion, the trapped Fermi SF has a
universal behavior not only in the BCS and unitarity limits but also along
the BCS-unitarity crossover. This allows a prediction of the static
properties   of a large
Fermi
SF in the crossover region from a
knowledge of the same of a small system through a  scaling
relation, cf. Eq. (\ref{scaling}).
The predicted energies of a Fermi SF  with $10^5$ atoms from a
knowledge of the same with 4   (10) atoms is found to have an
error of less than 3$\%$.
Although we used a DF equation in our study, the observed scaling(s)  
should be independent of the use of the DF equation and also 
of the LDA. The scaling(s) is a 
consequence of the dominance of the Pauli principle leading to a 
repulsive interaction, where the actual physical attraction plays a 
minor passive role.

In a Bose-Einstein condensate (BEC), there is no Pauli repulsion, and 
the physical interaction plays a major active role. Consequently, in a BEC, the 
stationary observables are very sensitive to the atomic interaction and 
no scaling should exist in a stationary observable as a function of $N$ 
from weak-coupling to unitarity crossover. Nevertheless, there have been 
systematic studies of static properties of a BEC as a function of $N$ 
\cite{yukalov}.  Fermi-Bose mapping, fermionization, and other considerations 
 of a 
strongly-interacting one-dimensional Bose gas are taken up   in Ref. 
\cite{yukalov2}.


The work was partially supported by the FAPESP and CNPq of Brazil.

\end{document}